\documentclass{emulateapj}

\begin{document}
\title{Composition of the Innermost Core Collapse Supernova Ejecta}
\author{C. Fr\"ohlich\altaffilmark{1},
  P. Hauser\altaffilmark{1},
  M.\ Liebend\"orfer\altaffilmark{2},
  G.\ Mart\'{\i}nez-Pinedo\altaffilmark{3},
  F.-K.\ Thielemann\altaffilmark{1},
  E.\ Bravo\altaffilmark{4},
  N.\ T.\ Zinner\altaffilmark{5},
  W.R.\ Hix\altaffilmark{6},
  K.\ Langanke\altaffilmark{7},
  A.\ Mezzacappa\altaffilmark{6},
  K.\ Nomoto\altaffilmark{8}%
}

\altaffiltext{1}{Departement f\"ur Physik und Astronomie, Universit\"at
  Basel, CH-4056 Basel, Switzerland}

\altaffiltext{2}{Canadian Institute for Theoretical Astrophysics,
  University of Toronto, Toronto ON M5S 3H8, Canada}

\altaffiltext{3}{ICREA and Institut d'Estudis Espacials de Catalunya,
  Universitat Aut\`onoma de Barcelona,
  E-08193 Barcelona, Spain}

\altaffiltext{4}{Universitat Polit\`ecnica de Catalunya,
  E-08034 Barcelona, Spain}

\altaffiltext{5}{Institute of Physics and Astronomy,
  Aarhus University, Aarhus C, Denmark}

\altaffiltext{6}{Physics Division, Oak Ridge National Laboratory,
  Oak Ridge TN 37831-6374, USA}

\altaffiltext{7}{Gesellschaft f\"ur Schwerionenforschung, D-64291 Darmstadt,
  Germany}

\altaffiltext{8}{Department of Astronomy, University of Tokyo,
  Tokyo 113-033, Japan}

\begin{abstract}

With presently known input physics and computer simulations in 1D, 
a self-consistent treatment of core collapse supernovae does not yet lead 
to successful explosions, while 2D models show some promise. Thus, 
there are strong indications
that the delayed neutrino mechanism works combined with a multi-D convection
treatment for unstable layers (possibly with the aid of rotation, magnetic
fields and/or still existent uncertainties in neutrino opacities).
On the other hand there is
a need to provide correct nucleosynthesis abundances
for the progressing field of galactic evolution and observations of low
metallicity stars. The innermost ejecta is directly affected
by the explosion mechanism, i.e.\ most strongly the yields of Fe-group nuclei
for which an induced piston or thermal bomb treatment will not provide the
correct yields because the effect of neutrino interactions is not included.
We apply parameterized variations to the neutrino scattering cross
sections in order to mimic in 1D the possible increase of neutrino luminosities
caused by uncertainties in proto-neutron star convection.
Alternatively, parameterized variations are applied to the
neutrino absorption cross sections on nucleons in the ``gain region''
to mimic the increase in neutrino energy deposition enabled by
convective turnover.
We find that both measures lead to similar results, causing explosions
and a $Y_e>0.5$ in the innermost ejected layers, due to the combined effect
of a short weak interaction time scale and a negligible electron
degeneracy, unveiling the proton-neutron mass difference.
We include all weak interactions (electron and positron capture, beta-decay, 
neutrino and antineutrino capture on nuclei, and neutrino
and antineutrino capture on nucleons) and present first nucleosynthesis
results for these innermost ejected layers to discuss how they improve
predictions for Fe-group nuclei.
The proton-rich environment results in enhanced abundances of $^{45}$Sc, 
$^{49}$Ti, and $^{64}$Zn as requested by chemical evolution studies and observations
of low metallicity stars
as well as appreciable production of nuclei in the mass range up to $A=80$.

\end{abstract}
\keywords{supernovae, nucleosynthesis}

\setcounter{footnote}{0}

\section{Introduction}
\label{sec:introduction}

The problem of core collapse supernova explosions is an old one and
the attempts to understand the mechanism have been ongoing for more than
30 years. The idea that a massive star proceeds through all burning
stages from H to Si-burning, finally leading to the collapse of the
resulting Fe-core to nuclear densities with formation of a
neutron star has long been discussed 
\citep[e.g.][]{Baade.Zwicky:1934,Oppenheimer.Snyder:1939,Arnett.Schramm:1973}.
Since the 1960s the explosion mechanism has been 
related to neutrino emission from the hot collapsed core
\citep[e.g.][]{Colgate.White:1966,Bethe.Wilson:1985,Bethe:1990}
interrupted by a
period when it was speculated that the strength of the bounce at
nuclear densities could permit shock waves with sufficient energies to
lead to prompt explosions 
\citep[e.g.][]{Baron.Cooperstein.Kahana:1985,Baron.Bethe.ea:1987}.
However, introduction of previously
neglected neutrino scattering processes were introduced (e.g.
neutrino-electron scattering), 
which permitted the replacement of lost low energy neutrinos, led to a
continuous energy leakage and to the death
of the prompt shock within 10 ms after bounce 
\citep[]{Bruenn:1989,Myra.Bludman:1989}.

Since then, and with the first neutrino detection from a core collapse
supernova 
\cite[SN1987A, see e.g.][]{Koshiba:1992,Burrows:1990},
the hope has been that further improvements would lead to successful explosions 
via energy deposition through neutrino and antineutrino captures on
neutrons and protons ($\nu_e + n \rightarrow p + e^-$, $\bar\nu_e + p
\rightarrow n + e^+$). 
Two different paths were explored.
1.\ Convective instabilities, but with still simplified neutrino transport, 
causing either (a) convective transport in the core and 
leading to higher neutrino luminosities \citep[e.g.][]{Keil.Janka.Mueller:1996}
or (b) higher energy deposition efficiencies in convective regions
\citep{Mayle.Wilson:1988,Herant.Benz.ea:1994,Janka.Mueller:1996,%
Fryer.Warren:2004}.
2.\ Improved neutrino transport schemes, leading to 
higher neutrino luminosities via the full solution of the
Boltzmann transport equation for neutrino scattering and neutrino reactions 
\citep{Mezzacappa.Bruenn:1993a,Mezzacappa.Bruenn:1993b,Messer.Mezzacappa.ea:1998}.

However, in recent years 1D spherically and 2D rotationally symmetric 
radiation-hydro calculations have not yet shown successful
supernova explosions with the present knowledge of physical processes.
\citep{Rampp.Janka:2000,Mezzacappa.Liebendoerfer.ea:2001,%
Liebendoerfer.Mezzacappa.ea:2001a,Liebendoerfer.Mezzacappa.ea:2001b,%
Buras.Rampp.ea:2003,Janka.Buras.Rampp:2003,Hix.Messer.ea:2003,%
Langanke.Martinez-Pinedo.ea:2003,Thompson.Burrows.Pinto:2003}.
A very recent simulation of a 11.2~M$_{\odot}$ core collapse
shows a chance for successful weak explosions in a multi-D 
treatment with spectral neutrino transport \citep{Janka.Buras.ea:2005}.
This leaves us with two dilemmata. 
First, the fundamental problem
that the supernova mechanism is still not understood.
Second, there seems no way to predict the correct supernova nucleosynthesis
yields. This is a problem 
in itself, but is also a limitation for the rapidly
expanding field of galactic chemical evolution,
which is 
being energized 
by the large amount of recent abundance observations from low metallicity
stars \citep[e.g.][]{Argast.Samland.ea:2002,Argast.Samland.ea:2004,%
Sneden.Cowan.ea:2003,Cayrel.Depagne.ea:2004,Honda.Aoki.ea:2004,%
Frebel.ea:2005}.

Supernova nucleosynthesis predictions have a long tradition
\citep[]{Woosley.Weaver:1986,Thielemann.Hashimoto.Nomoto:1990,%
Woosley.Weaver:1995,Thielemann.Nomoto.Hashimoto:1996,%
Nomoto.Hashimoto.ea:1997,%
Nakamura.Umeda.ea:2001,Rauscher.Heger.ea:2002,%
Chieffi.Limongi:2002a,%
Chieffi.Limongi:2004,Umeda.Nomoto:2005}.
But all of these predictions relied on an artificially introduced 
explosion, either via a piston or a thermal bomb
\citep[]{Aufderheide.Baron.Thielemann:1991}
introduced into the progenitor star model. 
The mass cut between the ejecta and the remnant
does not emerge from the simulations, but has to be determined
from additional conditions.
While the usage of artificially introduced explosions is justifiable
for the outer stellar layers, provided we
know the correct explosion energy to be dumped into the shock front (on
the order of 10$^{51}$ erg seen in observations), it clearly is
incorrect for the innermost ejected layers which should be directly
related to the physical processes causing the explosion. 
This affects the Fe-group composition, discussed in detail in 
\citet{Thielemann.Nomoto.Hashimoto:1996},
hereinafter TNH96,
and \citet{Nakamura.Umeda.ea:1999},
which was also recognized as a clear problem by
\citet{Chieffi.Limongi:2002a} 
and \citet{Umeda.Nomoto:2002}.
The problem is also linked to the
so-called neutrino wind, emitted seconds after the supernova
explosion, and considered as a possible source of the \emph{r}-process to
produce the heaviest elements via neutron captures 
\citep{Takahashi.Witti.Janka:1994,Woosley.Wilson.ea:1994,Qian.Haxton.ea:1997,%
Thompson.Burrows.Meyer:2001,Wanajo.Kajino.ea:2001,Terasawa.Sumiyoshi.ea:2002,%
Thompson.Burrows.Pinto:2003}.

An indispensable quantity to correctly describe nucleosynthesis
in the innermost ejecta is the electron fraction
$Y_e=\langle Z / A\rangle$ 
in the layers undergoing explosive 
Si-burning. This $Y_e$ is set by the weak interactions in the explosively
burning layers, i.e.\ electron and positron captures, beta-decays, and
neutrino or antineutrino captures. The dominant reactions in hot
photodisintegrated matter, consisting mainly of neutrons and protons, are

\begin{displaymath}
  \begin{array}{l}
    \nu_e + n \rightleftarrows p + e^-\\ 
    \bar\nu_e + p \rightleftarrows n + e^+ .
  \end{array}
\end{displaymath}

In section \ref{sec:masscut} we will show that these reactions lead to 
significant changes in $Y_e$ toward equilibrium before the matter is ejected.
We will also
show that the resulting $Y_e$ in the innermost ejected layers is
close to 0.5, in some areas even exceeding 0.5. This has been strongly
postulated as a requirement in order not to violate abundance constraints
from galactic evolution and solar abundances (TNH96).

The question arises how one could realistically simulate this
behavior, given the existing problems with self-consistent explosions.
Discussed improvements which could lead to successful supernova
explosions are rotation and magnetic fields 
\citep[e.g.][]{Thompson:2000,Thompson.Quataert.Burrows:2004} or 
uncertainties in neutrino opacities 
\citep[see e.g.][]{Burrows.Reddy.Thompson:2004} 
or other microphysics properties. 
They would introduce
additional mixing at the neutrino sphere and convective transport 
or change the neutrino luminosity via improved opacities.
This indicates
the two options for successful explosion as discussed above: 
(a) enhanced neutrino luminosities or (b) enhanced deposition efficiencies 
for neutrino capture in convective layers. These effects can be simulated 
in two ways:
(a) Boosting the neutrino luminosity via a scaling (reduction) of the neutrino
scattering cross sections on nucleons while keeping the electron/positron and
neutrino/antineutrino capture cross sections on neutrons and protons
at their original values. 
(b) Boosting the energy deposition efficiencies by enhancing the neutrino 
and antineutrino captures on neutrons and protons.
Neither approachrepresents
a self-consistent treatment, but no external energy 
is required to produce (i) a successful explosion
with (ii) a consistently emerging mass cut between neutron star and ejecta.
Moreover, our treatment guarantees that $Y_e$ is consistently determined 
by all weak interaction processes.

\section{Matter in a neutrino field}

\label{sec:highye}
Even though it is uncertain
how significantly absorptions
of the neutrinos emitted from the protoneutron star surface contribute
to the revival of the shock, it is
necessary to include the neutrinos and their copious interaction
with the matter in the vicinity of the protoneutron star. If the explosion
is launched such that the mass cut is directly determined by the early
dynamics of the explosion, this neutrino heated material will contribute
to the deepest layers of the ejecta. 
If fallback occurs after the initial explosion, 
contributions 
by these innermost layers
are only possible 
if strong mixing occures.
In any case, this neutrino heated material will have significantly
changed its composition with respect to its original progenitor composition.
Hence, we investigate in this section the conditions in ejecta that
are subject to large neutrino fluxes.

The energy spectrum of the neutrinos is set
in the vicinity of the neutrino spheres
at a radius of initially \( \sim 70 \) km. Before the launch of the explosion,
about two thirds of the emitted neutrinos stem from the infalling
matter which is squeezed in the gravitational potential and settles
on the surface of the protoneutron star. 
The rise in the electron energies
by the compression leads to more electron neutrino
emissions than electron antineutrino emissions.
After the launch of the explosion,
this contribution will decrease with the accretion rate and the less 
accretion-sensitive neutrino diffusion flux from the hot protoneutron star 
will dominate. 
An energy equipartition among the different neutrino
flavors is expected to set in when the accretion
luminosities have reduced to a negligible contribution.
A more detailed description of this transition, however, 
requires multi-dimensional simulations
because the evolution of the accretion rate shows quite aspherical features
with narrow downflows and broad upflows \citep{Herant.Benz.ea:1994,%
Burrows.Hayes.Fryxell:1995,Janka.Mueller:1996,Buras.Rampp.ea:2003}
that are ignored in spherical symmetry. 
The emitted electron flavor
neutrinos may essentially interact with the material behind the accretion
shock out to radii of about \( 300 \) km ({}``essential'' meaning
electron fraction changes on a time scale of \( 100 \) ms).
The interactions decrease steeply with increasing radius 
due to geometric  $1/r^2$ dilution of the neutrino field.

In order to illustrate the basic behaviour of shock-heated matter
in a neutrino bath, we first consider only the four 
dominant reactions,
electron capture on free protons \( e^{-}+p\rightleftarrows n+\nu _{e} \),
and positron capture on free neutrons 
\( e^{+}+n\rightleftarrows p+\bar{\nu }_{e} \),
and their inverse reactions. Two independent conditions are required
to specify the electron fraction and the entropy of the material,
for example weak equilibrium and balance in the energy exchange
with neutrinos.

The change of the electron fraction, \( Y_{e} \), with time, \( t \),
is given by Eqs.\ (C15) and (C20) in \citet{Bruenn:1985}. The neutrino
opacities, \( \chi  \), and emissivities, \( j \), are linked by
the reciprocity relation 
(detailed balance)
described in Eqs.\ (C7) and (C8) in the above reference.
The reciprocity relation involves the temperature, \( kT=\beta ^{-1} \),
the neutrino energy, \( E \), measured in the rest frame of the fluid,
and the chemical potentials, \( \mu _{n} \), \( \mu _{p} \), and
\( \mu _{e} \), for neutrons, protons, and electrons respectively.
We can therefore label contributions from electron,
positron, neutrino, and antineutrino capture with \( EC \), \( PC \),
\( NC \), and \( AC \) respectively, and express the opacities in
\( NC \) and \( AC \) by the neutrino emissivities. After having
collected all terms that do not depend on the neutrino energy into
a common factor, \( K \), we write the total change in the electron
fraction in the following form:
\begin{eqnarray}
\frac{1}{c}\frac{dY_{e}}{dt} & = & K\int dEE^{2} 
\left[ h(E+Q)\left( -EC+NC\right) \right. \nonumber \\
 &+& \left. \Theta \left( E-Q-m_{e}\right) h(E-Q)\left( PC-AC\right) \right] .
\label{ye_evolution} 
\end{eqnarray}
Here, the details of the roughly quadratic energy dependence of the
cross sections are hidden in the function\[
h(x)=x^{2}\left[ 1-\left( \frac{m_{e}c^{2}}{x}\right) ^{2}\right] ^{1/2},\]
and a step function \( \Theta (x) \) is used to describe the energy
threshold in the positron and antineutrino capture reactions. A very
similar equation can be used to describe the change of the specific
internal energy, \( e \), of the fluid due to neutrino interactions:
\begin{eqnarray}
\frac{1}{c}\frac{de}{dt} & = & K\int dEE^{3}
\left[ h(E+Q)\left( -EC+NC\right) \right. \nonumber \\
 &+& \left. \Theta \left( E-Q-m_{e}\right) h(E-Q)\left( -PC+AC\right) \right] .
\label{e_evolution} 
\end{eqnarray}
The density in the supernova ejecta is low enough that we
can neglect the nucleon degeneracy and nucleon final state
blocking described in Eq. (C14) of \citet{Bruenn:1985}.
The contributions to Eqs.\ (\ref{ye_evolution}) and (\ref{e_evolution})
from the individual reactions are given by
\begin{eqnarray}
EC & = & \frac{1}{1+e^{\beta \left( E+Q-\mu _{e}\right) }}n_{p}\left( 1-f_{\nu
}\right) \nonumber \\
NC & = & \frac{e^{\beta \left( E+Q-\mu _{e}\right) }}{1+e^{\beta \left(
E+Q-\mu _{e}\right) }}n_{n}f_{\nu }\nonumber \\
PC & = & \frac{1}{1+e^{\beta \left( E-Q+\mu _{e}\right) }}n_{n}\left(
1-f_{\bar{\nu }}\right) \nonumber \\
AC & = & \frac{e^{\beta \left( E-Q+\mu _{e}\right) }}{1+e^{\beta \left(
E-Q+\mu _{e}\right) }}n_{p}f_{\bar{\nu }}\label{eq_interactions}
\end{eqnarray}
where \(n_{n}\) and \(n_{p}\) are the neutron and proton number densities,
respectively, and \( f(E) \) is the neutrino distribution function in the rest
frame of the fluid.

Note that Eqs.\ (\ref{ye_evolution}) and (\ref{e_evolution}) do not
presume that the neutrinos are in equilibrium with matter, nor that
they assume any particular spectrum. Some of these assumptions, however,
lead to useful analytical formulas for the equilibrium electron fraction.
Dominance of the neutrino absorption terms has been assumed for the 
investigation of the \emph{r}-process in the neutrino wind of a 
protoneutron star 
\citep{Qian.Woosley:1996, Thompson.Burrows.Meyer:2001};
and the cases where emission terms dominate or where the neutrinos
are in thermal equilibrium have been analysed in a study of gamma-ray
burst fireballs \citep{Beloborodov:2003}. The balance between the
four reactions in Eq.\ (\ref{eq_interactions}) is determined by 
the ratio of the neutron and proton number densities and the exponential
\( \exp (\beta [E\pm Q\mp \mu _{e}]) \).
Its energy integral depends on a competition between the
neutrino energy, \( E \) (in NC and AC populated according to the
neutrino distribution functions \( f \)),
the mass difference between neutrons and
protons, \( Q \), and the electron chemical potential, \( \mu _{e} \).
Depending on the conditions, any
one of these three quantities can 
assume a dominant influence on the balance in above reactions.

For neutrinos at high energies
compared to $|Q-\mu_e|$
the term \( \exp (\beta E) \)
is large and the neutrino absorption terms, \( NC \) and \( AC \)
in Eq.\ (\ref{eq_interactions}), dominate over neutrino emission.
Hence, if the abundance of these high energy neutrinos is large, 
the equilibrium \( Y_{e} \) is determined by the balance between 
neutrino and antineutrino absorption and therefore dependent 
on the unknown neutrino distribution functions. 
The rate of change of $Y_e$ is given by
\begin{equation}
\frac{dY_e}{dt} \approx \lambda_{\nu_e n} -
                  Y_e (\lambda_{\nu_e n} + \lambda_{\overline{\nu}_ep}) .
\end{equation}
Using Eqs.\ (64a) and (64b) of \citet{Qian.Woosley:1996} for $\lambda_{\nu_e n}$
and $\lambda_{\overline{\nu}_ep}$ and using the fact that neutrino
and antineutrino luminosities are similar
(upper right panel of Figure \ref{fig:zone5})
it can be shown that $Y_e>0.5$ is achieved provided that
$4(m_n -m_p) > (\varepsilon_{\overline{\nu}}-\varepsilon_{\nu})$\
(see also Fig.\ 5 in \citet{Qian.Woosley:1996} and the discussion
in \( \S 3 \) of \citet{Hoffman.Woosley.ea:1996}).

However, at earlier time, and as long as the
eventual ejecta are close to the neutron star, all four reactions
in Eq.\ (\ref{eq_interactions}) are active and the neutrino distribution
functions are nontrivial functions of the accretion rate, the distance
from the neutrino spheres, and the local weak interactions. Changes in
the electron fraction are not only determined by the local neutrino fluxes
and spectra, but also by the ability of the matter
to accept captured electrons or neutrinos at the given conditions.
Especially, when the electrons are degenerate
the electron chemical potential can dominate the exponential for
average neutrino energies.
In this case, \( \exp \left( \beta \left( E+Q-\mu _{e}\right) \right)  \)
in Eq.\ (\ref{eq_interactions}) is small and \( \exp \left( \beta \left( E-Q+\mu _{e}\right) \right)  \)
becomes large. Hence, neutrino absorption, \( NC \), and positron
capture, \( PC \), are suppressed and the electron fraction decreases
because of more prolific electron captures and antineutrino absorptions.
Balance is only established when the ratio between proton and neutron
number densities has sufficiently decreased to compensate for the suppression
of \( NC \) and \( PC \) introduced by the exponential. This leads to
\( n_{p} < n_{n} \) and an equilibrium electron fraction \( Y_{e}<0.5 \).

Finally, in a plasma with nondegenerate electrons,
the electron chemical potential becomes rather small
so that the neutron to proton mass difference,
\( Q \), may actually dominate the exponentials in the balance equations,
making \( \exp \left( \beta \left( E+Q-\mu _{e}\right) \right)  \)
in Eq.\ (\ref{eq_interactions}) larger and \( \exp \left( \beta \left( E-Q+\mu \
_{e}\right) \right)  \)
smaller.
Reversing the trend
under degenerate conditions, \( NC \)
and \( PC \) are favored and \( n_{p} > n_{n} \)
is assumed in equilibrium. For not too different neutrino
and antineutrino fluxes and spectra, equilibrium will
establish at \( Y_{e}>0.5 \).
According to the analytical investigation
in \citet{Beloborodov:2003} for the \( EC \) and \( PC \) reactions,
this situation will occur if the electron chemical potential fulfills
the condition \( \mu _{e}<Q/2 \). The larger binding energy favors
protons over neutrons. High electron fractions below
\( Y_{e}=0.5 \) have been predicted for supernova explosions 
\citep{Fuller.Meyer:1995,Thompson:2000}.
But recent supernova simulations with accurate neutrino transport
have even exceeded the estimates, consistently finding values of \( Y_{e}>0.5 \)
in the vicinity of the mass cut in explosion settings 
\citep{Liebendoerfer.Mezzacappa.ea:2001a,%
Buras.Rampp.ea:2003,Janka.Buras.Rampp:2003,%
Thompson.Quataert.Burrows:2004,Pruet.Woosley.ea:2004,Frohlich.Hauser.ea:2005}.

\section{Hydrodynamical simulations}

\label{sec:hydro}
The framework for this investigation are spherically
symmetric simulations with implicit general relativistic Boltzmann neutrino
transport, 
see \citet{Mezzacappa.Messer:1999}
and \cite{Liebendoerfer.Messer.ea:2004}
for a detailed description of the code \textsc{agile-boltztran}. It
features a dynamically adaptive grid 
\citep{Liebendoerfer.Rosswog.Thielemann:2002} that 
concentrates grid points at
the developing mass cut. The simulations
are performed until 
the
density 
drops to
\( \sim 10^{6} \) g/cm\( ^{3} \)
in the region of bifurcation between the ejecta and the
remnant. At this time, the mass contained in radial mass zones 
is becoming very small and the run experiences
ill-conditioned Jacobian matrices
in the Newton-Raphson scheme.
The simulations are then continued by an explicit
hydrodynamic code
until the temperature 
falls
below $T = 2 \times 10^8$ K.
This code employs an explicit difference scheme similar to
\citet*[][]{Colgate.White:1966} 
and a simplified nuclear reaction network
as explained in \citet{Bravo.Dominguez.ea:1993}.

We use two different approaches to enforce explosions in otherwise
non-explosive supernova models. We parametrize the neutral current neutrino 
scattering opacities on free nucleons with a factor ranging from 0.1
to 0.7 and use a finite differencing%
\footnote{According to Eq.\ (91) in \citep{Mezzacappa.Bruenn:1993a} instead of
Eq.\ (56) in \citep{Liebendoerfer.Messer.ea:2004}, see sections 3.3.2 and
4.1 in the latter reference for details.
} that helps to artificially increase the diffusive fluxes in regions
of very high matter density. The net result is a faster deleptonization
of the protoneutron star such that the neutrino luminosities are boosted
in the heating region. For the sake of computational efficiency, this
first series of parametrized runs (series A) has been calculated with
the lowest possible angular resolution, involving only inwards and
outwards propagating neutrinos. However, all of these measures only
affect the propagation of neutrinos in the model; the models are still
closed and respect energy and lepton number conservation.
We expect that series A represents a simplification
of the phenomenology of supernovae that would be driven by higher
neutrino luminosities than in the standard 
cases, for example
different forms of protoneutron star convection 
\citep{Wilson.Mayle:1993, Keil.Janka.Mueller:1996, Mezzacappa.Calder.ea:1998, Bruenn.Raley.Mezzacappa:2004}
or improvements in the uncertain 
nuclear matter physics.

With progress in computer speed and code parallelization, we were
able to perform simulations using standard resolution (6 angular bins,
12 energy groups) for the Boltzmann neutrino transport in the parameter
study for series B. Series B also includes the weak magnetism corrections
in the neutrino cross sections \citep{Horowitz:2002}. Explosions are
enforced by multiplying the absorptivities  and emissivities (i.e.\ the 
reaction rates for forward and backward reactions in 
$\nu_e + n \rightleftarrows p + e^-$ and 
$\bar\nu_e + p \rightleftarrows n + e^+$) 
in the heating region by equal factors. This reduces the time scale for
neutrino heating without changing the important equilibrium \( Y_{e} \)
and temperature. We hope to mimic with this approach a potentially
increased heating efficiency in the heating region as it is expected
in combination with overturn in this convectively unstable domain
\citep{Herant.Benz.ea:1994,Burrows.Hayes.Fryxell:1995,Janka.Mueller:1996,Mezzacappa.Calder.ea:1998,Buras.Rampp.ea:2003}.

\begin{table}

\caption{Name and properties of the discussed runs}

\centering

\begin{tabular}{cccccc}
\tableline
\tableline
Run&
 Parameter&
 \( E_{\mathrm{expl}} \) {[}erg{]}&
 \( m_{\mathrm{cut}} \){[}M\( _{\odot } \){]}&
 \( t_{v>0} \){[}s{]}&
 \( t_{\mathrm{end}} \){[}s{]}\\
\tableline
A60&
 60\% scatt.&
 \( 0.24\times 10^{51} \)&
 \( 1.585 \)&
 \( 0.46 \)&
 \( 0.64 \)\\
 A40&
 40\% scatt.&
 \( 0.78\times 10^{51} \)&
 \( 1.511 \)&
 \( 0.27 \)&
 \( 0.53 \)\\
 A20&
 20\% scatt.&
 \( 1.24\times 10^{51} \)&
 \( 1.444 \)&
 \( 0.20 \)&
 \( 0.44 \)\\
 B05&
 factor 5&
 \( 0.31\times 10^{51} \)&
 \( 1.586 \)&
 \( 0.38 \)&
 \( 0.53 \)\\
 B07&
 factor 7&
 \( 0.78\times 10^{51} \)&
 \( 1.531 \)&
 \( 0.26 \)&
 \( 0.43 \)\\
 B10&
 factor 10&
 \( 1.12\times 10^{51} \)&
 \( 1.509 \)&
 \( 0.24 \)&
 \( 0.40 \) \\
\tableline
\label{table_runs}
\end{tabular}
\tablecomments{The parameter of series
A specifies the percentage of neutral current interactions considered
in the model. The parameter of series B specifies the reduction
of the heating time scale. The time after bounce where we had to stop
the runs with neutrino transport is displayed in the last column labelled
by \( t_{\mathrm{end}} \). The time of the first appearance
of positive velocities is given in the column \( t_{v>0} \).
The mass cut \( m_{\mathrm{cut}} \) has
been determined at the point where the total energy integrated from
outside inwards reaches a maximum.
The estimate for the explosion energy \( E_{\mathrm{expl}} \) has been
composed from the total energy of the unbound material between the masscut
and the shock front at \( t_{\mathrm{end}} \) (mostly material that
was in NSE) and a correction for the total energy of the bound
layers ahead of the shock at progenitor composition.}

\end{table}

All models are based on a progenitor model with a main sequence mass
of \( 20 \) M\( _{\odot } \) \citep{Nomoto.Hashimoto:1988}. The
parameters in series A and B are chosen such that each series contributes
with a barely exploding model, an extremely exploding model (in terms
of parameter range, the explosion energy itself seems to saturate
around \( 1.2\times 10^{51} \) erg), and a model with average parameter
setting. Important properties of the different runs are listed in
Table \ref{table_runs}.

\begin{figure*}
{\centering \resizebox*{1\textwidth}{!}{\includegraphics{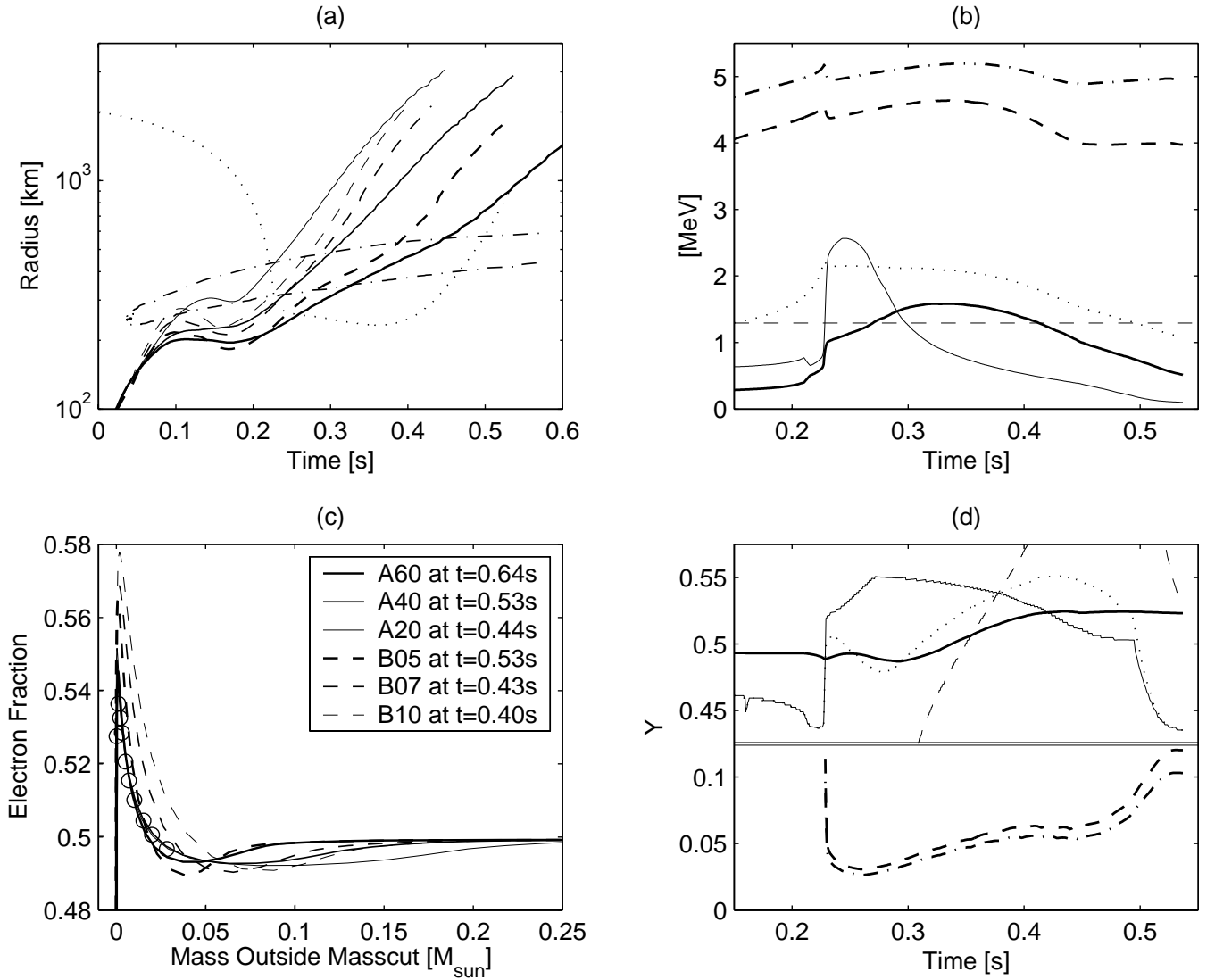}} \par}

\caption{(a) Trajectories of the (accretion-) shock as a function of time.
Series A (solid lines) explodes because of artificially increased core
diffusion luminosities. The neutral current neutrino opacity is used
as parameter with \protect\( 20 \protect\)\% of the standard values
(thin line), with \protect\( 40 \protect\)\% (medium line width), and with
\protect\( 60 \protect\)\% (thick line).
Series B (dashed lines) explodes because of artificially accelerated
neutrino absorption and emission processes in the heating region. The acceleration
factors are 10 (thin line), 7 (medium line width), and 5 (thick line).
See also the legend in panel (c) for an identification of the
shock trajectories. The dotted line traces the
trajectory of a representative fluid element located at \protect\( 0.005\protect
 \)
M\protect\( _{\odot }\protect \) outside of the estimated mass cut.
Its electron fraction and specific energy can change by weak interactions
on a time scale represented by the dot-dashed lines (reaction time
scale as a function of radius). The lower branch
belongs to the infall phase and the upper branch to the ejection.
\protect \\
(b) Energy scales sampled at the representative fluid element as a function
of time. Thick lines show the fluid temperature (solid), the neutrino temperature
(dashed), and the antineutrino temperature (dot-dashed). Thin lines
represent the electron chemical potential (solid), the mass difference
between neutron and proton (dashed), and the temperature that would be
obtained after infinite exposure to the neutrino field (dotted). The
electrons in the representative fluid element are degenerate during
collapse and after shock compression. Neutrino heating and expansion
during the ejection lifts the electron degeneracy and the neutron to
proton mass difference becomes the dominant energy scale.
\protect \\
(c) Electron fraction as function of mass when the runs are stopped
(\protect\( 0.4-0.6\protect \) s after bounce) with (A) reduction
of the neutral current scattering opacities
(solid lines), or (B) enhancement of the reaction timescales (dashed
lines). The open circles represent the final electron fractions of
run A40 after the freezout of charged current interactions. 
\label{fig:evol}
\protect \\
(d) Abundances sampled at the representative fluid element. Thick lines
show the electron
fraction (solid), neutrino fraction (dashed), and antineutrino fraction
(dot-dashed). Thin lines represent electron fractions that would obtain
after infinite exposure to the neutrino field. Only neutrino absorptions
have been considered for the solid line and only emissions have been
considered for the dashed line. The dotted line includes all charged
current reactions. Panel (b) and (d) show that the electron fraction
is first kept high by neutrino absorptions, later by neutrino emissions.
The electron fraction at freezout is determined by competition between
the neutrino interaction rates and the matter ejection timescale.
}
\end{figure*}

Figure \ref{fig:evol}a presents an overview of the shock trajectories.
Runs from series A are presented with solid lines and runs from series
B with dashed lines. The legend in Figure \ref{fig:evol}c also applies
to Figure \ref{fig:evol}a. In all runs, the accretion front stalls
at about \( 100 \) ms after bounce at a radius between \( 180 \)
and \( 300 \) km, depending on the parameters. The accretion front
is slowly receding in the more optimistic models. Shortly before \( 200 \)
ms after bounce, the accretion front moves outward again. There may
still be some additional delay until the inwards drifting material
behind the shock reverses its velocity and starts to accumulate kinetic
energy for the ejection. This happens at \( 199 \) ms after bounce
for the fastest explosion (A20) and at \( 461 \) ms after bounce
for the slowest run (A60). Bruenn's suggestion to locate the mass cut where
the integrated total energy of all external material assumes a maximum
agrees well with the actual bifurcation in the mass trajectories.
The mass cuts, \( m_{\mathrm{cut}} \), range from \( 1.444 \) M\( _{\odot } \)
to \( 1.585 \) M\( _{\odot } \).
Realistic 3D calculations where convection (responsible for the corrections
applied to the weak rates in both series A and B) turned on in a delayed 
fashion could delay the explosions and lead to larger mass cuts.

\section{Conditions of matter in the vicinity of the mass cut}

\label{sec:masscut}
In the following, we trace a mass element in the
exemplary run A40. We choose a mass element that is \( 0.005 \) M\( _{\odot } \)
outside of the mass cut. The trajectory of this mass element is represented
in Figure \ref{fig:evol}a by a dotted line: At first, the element is
falling into the gravitational potential. After \( 200 \) ms it
passes through the accretion shock at about \( 300 \) km radius and
is instantaneously decelerated. A second phase of drifting around
in the heating region follows until about \( 400 \) ms after bounce.
Finally, the mass element is ejected to larger radii.

Figure \ref{fig:evol}b illustrates important energy scales along the
trajectory. The dashed and dash-dotted thick lines at the top of the
graph indicate the neutrino temperature for the electron neutrinos
and antineutrinos respectively. They show a rising trend in the first
half of the graph. This is because the protoneutron star shrinks and
the neutrino spheres become hotter as they shift deeper into the gravitational
well. The discontinuity at the crossing of the shock front stems from
the Doppler shift when the mass element crosses the velocity jump
at the accretion front. The change of the rise into a decline around
\( t=350 \) ms after bounce is due to the decrease of the accretion
rate after the lauch of the explosion. Rising neutrino temperatures
are resumed at a very small accretion rate after \( t=450 \) ms.

With the full neutrino spectrum and abundances from the simulation
and with the matter density as input, we calculate the equilibrium matter
temperature along the trajectory according to Eq.\ (\ref{e_evolution})
by requiring \( de/dt=0 \). For consistency with the simulation,
we have also included the charged current reactions with nuclei according
to the simple model described in \citet{Bruenn:1985}. The dotted line
in Figure \ref{fig:evol}b shows the equilibrium temperature of matter
subject to the neutrino luminosities (the neutrinos themselves are
not in thermal equilibrium with matter, their temperature is set in
the vicinity of the neutrino spheres where the matter temperature
is higher). The lower part of Figure \ref{fig:evol}b shows the matter
temperature (thick solid line) and the electron chemical potential
(thin solid line). The electrons are degenerate in the cool infalling
matter. The first little blip in the trajectory after \( t=200 \)
ms is due to the burning of the initial silicon layer to nuclear statistical
equilibrium. 
It causes a slight rise in the temperature and decline
in the electron chemical potential. The pronounced step up in both
quantities is due to shock compression when the mass element hits
the accretion front. During the drift in the heating region, we note
a temperature increase towards temperature balance (dotted line) by
neutrino heating. The onset of the explosion during this time also
leads to an expansion and drop in matter density. Both effects work
together to lift the electron degeneracy shortly before \( 300 \)
ms after bounce (crossing of temperature and electron chemical potential
lines). The evolution during the third phase (ejection) is characterised
by a density decrease. The weak interaction rates decrease and the
temperature declines due to adiabatic expansion. The electrons stay
nondegenerate and the electron chemical potential remains smaller
than the neutron to proton mass difference (dashed thin line). In
contrast to the electron-degenerate conditions found in past supernova
simulations that fail to explode, the expanding hot plasma under neutrino
irradiation favors electron fractions that exceed \( 0.5 \) as discussed
in section \ref{sec:highye}.

The lower part in Figure \ref{fig:evol}d shows the neutrino and antineutrino
abundances with dashed and dash-dotted lines respectively. The variations
are due to density changes rather than luminosity variations. The
upper part of Figure \ref{fig:evol}d shows the electron fraction from
the simulation (thick solid line) and the equilibrium value determined
by Eq.\ (\ref{ye_evolution}) (dotted line). The dash-dotted line in
Figure \ref{fig:evol}a shows the reaction time scale as a function
of radius. The upper branch belongs to infall, the lower branch to
the ejection. Outside a radius of \( 600 \) km the reaction time
scale is much larger than the dynamical time scale; during the drift
phase of our mass element in the heating region it assumes values
around \( 50 \) ms. Thus, the low electron fraction during infall
is mainly set by the progenitor model. Before the shock front is crossed
by the mass trajectory, the equilibrium \( Y_{e} \) is also low because
many neutrons are bound in nuclei and not available as targets for
antineutrino absorption. After the shock transition, matter is dissociated
and higher electron fractions are favored. At first sight, the equilibrium
electron fraction appears higher than expected at the given electron
degeneracy. The reason are neutrino absorption rates that are by an order of
magnitude larger than the neutrino emission rates at these moderate temperatures.
The thin solid line shows the high electron fraction equilibrium as
it would evolve if only neutrino absorption were considered. The emission
reactions alone favor a much lower equilibrium \( Y_{e} \) (thin
dashed line) because there are only few positrons to capture under
degenerate conditions. With the following rise of the temperature,
however, the neutrino emission reactions (e.g.\ electron capture) 
gain weight with respect to the absorption
reactions and the equilibrium \( Y_{e} \) correspondingly adjusts
to lower values in the time window between \( t=235-275 \)~ms. But
as the electron degeneracy is lifted with further temperature increase
and expansion, and the electron chemical potential dips below half
the neutron to proton mass difference, the emission rates start 
to join the absorption rates in favoring higher electron fractions
(steep rise of the thin dashed line). The equilibrium \( Y_{e} \) 
increases again. The
descent at very late time is, as in the beginning, due to the reappearance
of nuclei. The electron fraction in the simulation (thick solid line)
can now easily be understood: At each time it evolves towards the
equilibrium value for the combined reactions (dotted line) at the
pace of the local reaction time scale. It freezes out when the mass
element is ejected. Note that for an analytical estimate of the electron
fraction in our application one would have to combine the approximations
for neutrino absorption rates in Eq.\ (64) in \citet{Qian.Woosley:1996}
with the approximation for neutrino emission rates in Eqs.\ (9-10)
in \citet{Beloborodov:2003} and to consider the reaction time scale
in order to find the correct freeze-out value in the \( Y_{e} \)
evolution.

We find that all simulations that lead to an explosion by neutrino
heating develop a proton-rich environment around the mass cut with
\( Y_{e}>0.5 \). This is illustrated by the electron fraction profiles
shown in Figure \ref{fig:evol}c. 
The open circles denote the final (i.e.\ at $T<2 \times 10^8$~K) 
electron fraction for the run A40.
The mass scale is normalized to the
respective mass cut. The different runs from series A show an almost
identical electron fraction profile at the mass cut. The competition
by the reaction and ejection time scale is not directly influenced
by the different explosion parameters, i.e. the enhanced neutrino
diffusion at higher densities. The electron fraction profiles of series
B, however, respond to the different reaction time scales set by the
explosion parameters in the heating region.

The electron fractions
around and outside of \( m_{\mathrm{cut}}+0.1 \)~M\( _{\odot } \) are still
close to the progenitor values. Differences in this region stem from
the different locations of the mass cuts within the progenitor composition.
It is important to note that the investigated region at the mass cut
is highly unstable against convection because of a large negative
entropy gradient. It is likely that the discrepancies in \( Y_{e} \)
are heavily mixed on a dynamical time scale \citep{Kifonidis.Plewa.ea:2003}.
We expect, however, that the \( Y_{e} \) remains high in an averaged
sense \citep[see also][]{Pruet.Woosley.ea:2004}. 
Moreover, matter blobs that leave the heating region in an environment
of large convective turnover may still show qualitatively similar
features in comparison with our spherically symmetric shells, because
the high electron fraction in the neutrino field is enabled by the
discussed general features of expanding hot matter. We believe that
the dependence on the details of our different simulations is small.

\section{Nucleosynthesis}
\label{sec:nucleosynthesis}

For the nucleosynthesis results presented here, we consider only the
first few zones outside of the mass cut enclosing a few hundredths of 
a solar mass where values of $Y_e$ higher than 0.5 are achieved.  
To determine the final electron fraction $Y_e$ in supernova ejecta 
it is necessary to include neutrino absorption reactions on neutrons
and on protons as well as electron and positron captures reactions.

\begin{figure*}
  \centering
{\centering \resizebox*{1\textwidth}{!}{\includegraphics[angle=0]{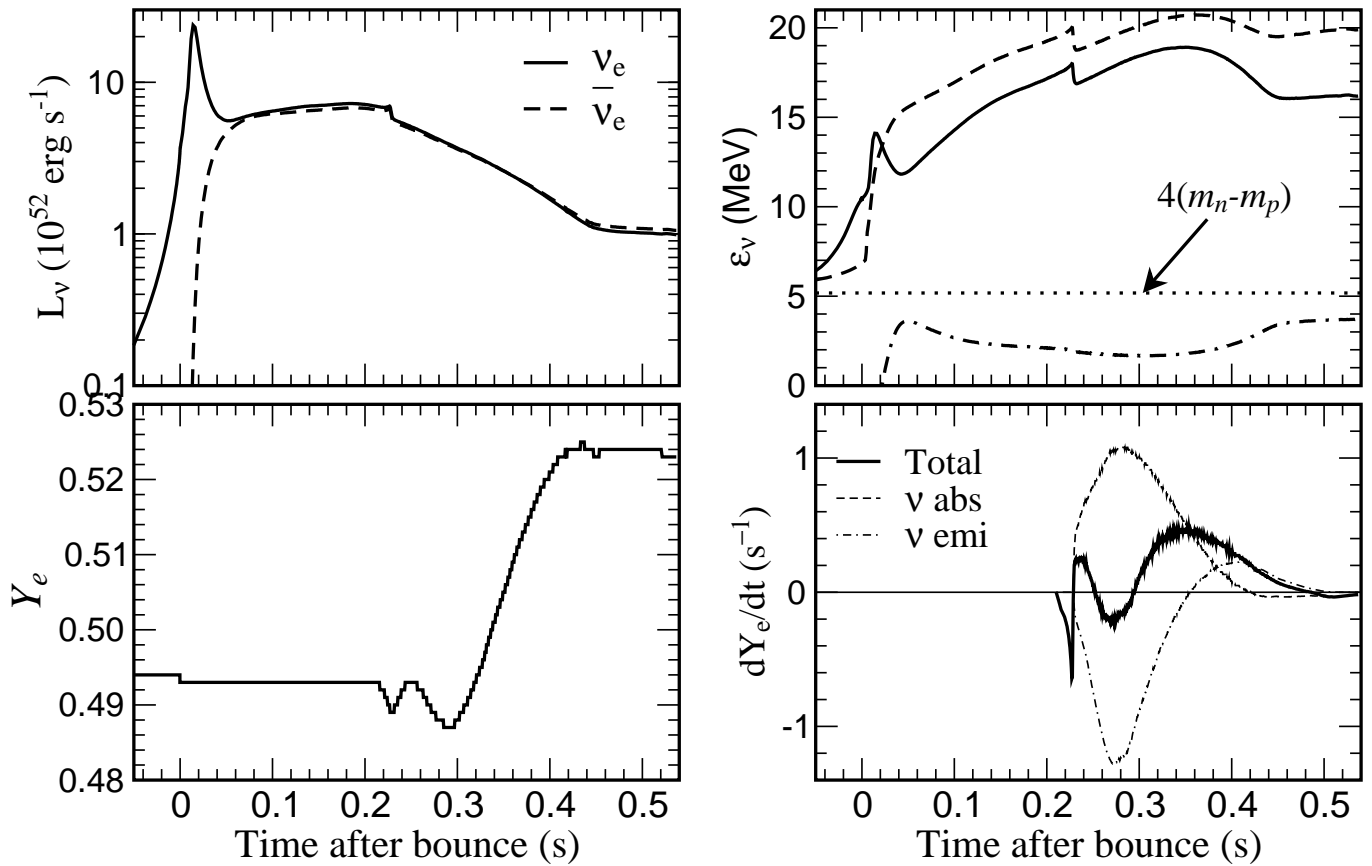}} \par}
  \caption{Time evolution after core bounce of an ejected layer at
           0.005~M$_{\odot}$ outside of the mass cut from a 20~M$_\odot$
           supernova progenitor.
           Top left: Luminosity of neutrinos and antineutrinos
                     felt by a Lagrangian mass zone.
           Bottom right: Electron fraction $Y_e$.
           Top left: Average neutrino energy 
                     $\varepsilon_{\nu}=\langle E^2_{\nu} \rangle / \langle E_{\nu} \rangle$ (thick solid and dashed lines).
                     The difference in average neutrino energy,
                     $(\varepsilon_{\overline{\nu}}-\varepsilon_{\nu})$
                     (thick dot-dashed line) and four times the mass
                     difference between neutron and protons (thin dashed line)
                     are shown in the lower part.
           Bottom right: Individual weak interaction contributions
                         leading to $Y_e>0.5$.
                     The individual contributions from neutrino/antineutrino
                     captures and electron/positron captures are a factor ten
                     larger than the total resulting $dY_e/dt$.
          }
  \label{fig:zone5}
\end{figure*}

An example of the influence of the individual weak interaction contributions
leading to $Y_e>0.5$ is given in Figure \ref{fig:zone5} (bottom right)
for an exploratory study of one mass zone.
Also shown are the neutrino luminosities $L_{\nu}$ (top left) and energies
$\varepsilon_{\nu} = \langle E^2_{\nu} \rangle / \langle E_{\nu} \rangle$
(top right).
For this exemplary mass zone it can be seen in the upper right
panel of Figure \ref{fig:zone5} that 
$(\varepsilon_{\overline{\nu}}-\varepsilon_{\nu})$ is always smaller than
$4(m_n -m_p)$.
There are several phases that can be identified in Figure \ref{fig:zone5}
and that have been discussed in section \ref{sec:highye}.
At early times ($t<0.3$~s) matter is degenerate and electron capture dominates.
At the same time matter is being heated by neutrino energy deposition and
around $t\thickapprox 0.3$~s the degeneracy is lifted (see upper panel of Figure
\ref{fig:zone5}). At this time, the ratio between electron captures and 
positron captures significantly decreases and
neutrino absorption reactions start to dominate
the change of $Y_e$ and, as discussed before, the average neutrino energies
favor $Y_e>0.5$. As the matter expands the density decreases, reducing the electron
chemical potential. This 
results in positron captures dominating electron captures beginning
around $t\thickapprox 0.3$~s.  From this time, the combined
effect of positron capture and $\nu_e$ absorption contributes to the
final increase of $Y_e$.

\begin{figure}
  \centering
\includegraphics[angle=0,width=\linewidth]{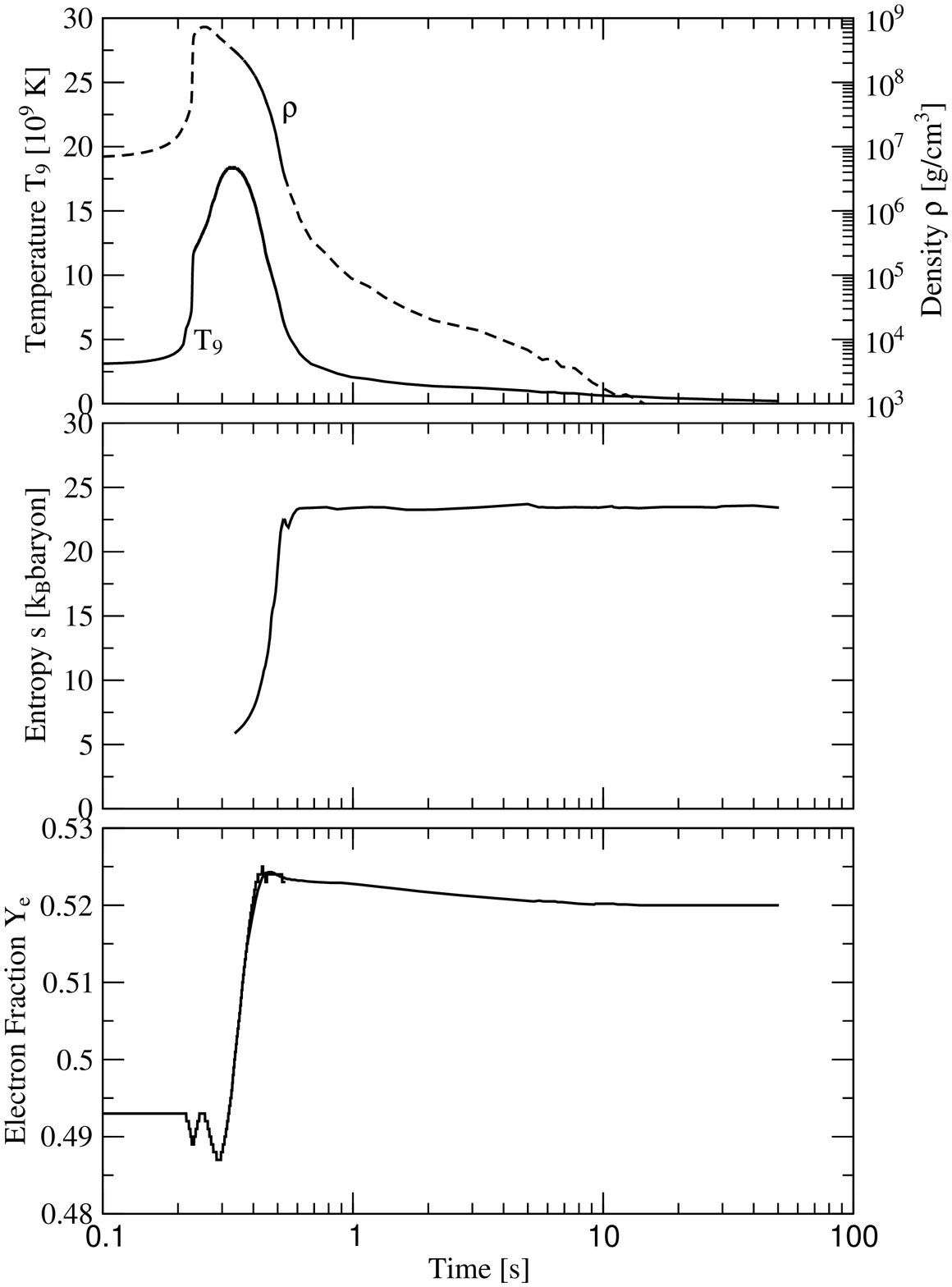}
  \caption{Evolution after core bounce for a representative ejected layer at
           0.005 M${_\odot}$ outside of the mass cut for model A40.
           For this layer the time for cooling from $T_9=2$ to
           $T_9=0.8$ is about 6 s.
           Top: Temperature (solid line) and density (dashed line)
           evolution.
           Middle: Entropy $s$ per baryon in units of the Boltzmann
           constant.
           Bottom: Electron fraction $Y_e$.
           The final decline is due to $\beta$-decays of the
           nucleosynthesis products.
           }
  \label{fig:longzone5}
\end{figure}

Figure \ref{fig:longzone5} shows the time evolution for a representative 
layer of models A40 (scattering cross sections on nucleons reduced by 40\%)
for the whole computational time. The final decline in the electron
fraction $Y_e$ is due to $\beta$-decays of the nucleosynthesis products.

The position of the mass cut emerges
consistently from the simulation as the region of bifurcation in which
the density has dropped below $\sim 10^6$  $\mathrm{g/cm^3}$. 
Based on the temperature-density profiles of all the matter in our hydrodynamical 
simulation, the detailed nucleosynthesis is calculated in a postprocessing 
framework
for the temperature range $T \ge 2 \times 10^8$ K.
The extended nuclear reaction network used consists of 1072
nuclei with $1\le Z\le50$,
see Table \ref{tab:network}.
Neutral and charged particle reactions are taken
from the recent REACLIB compilation 
(containing experimental rates by \citet{Angulo.Arnould.ea:1999} (NACRE)
and theoretical predictions by \citet{Rauscher.Thielemann:2000},
and are the same as used in \citet{Schatz.Aprahamian.ea:2001}).
For the weak interaction rates (electron/positron captures and beta decays) the
rates by \citet{Fuller.Fowler.Newman:1982a,Fuller.Fowler.Newman:1982b} 
are used for nuclei with $A\le45$ (\emph{sd}-shell). 
In the mass range $45< A \le 65$ (\emph{pf}-shell) the 
extended tabulation by \citet{Langanke.Martinez-Pinedo:2001} is used.
In Table \ref{tab:weaknetwork} a detailed list of nuclei is given
 for which the above weak rates were utilized.
The rates for neutrino and antineutrino captures on nuclei 
for the whole range of nuclei in the network 
are taken from a recent calculation \citep{Zinner.Langanke:2004},
based on the random phase approximation calculations of 
\citet{Langanke.Kolbe:2001,Langanke.Kolbe:2002}.
A complete list of nuclei for which neutrino and antineutrino capture reaction 
rates were included is shown in Table \ref{tab:neutrinonetwork}.
Hence, all weak interactions responsible for changes of $Y_e$ are taken 
into account in the reaction network,
namely: neutrino/antineutrino capture on free neutrons and protons, 
neutrino/antineutrino capture on nuclei, 
electron/positron capture, and $\beta^-$/$\beta^+$ decays.
Neutrino scattering processes 
do no contribute to abundance changes and are thus
not included in the reaction network 
used for postprocessing.
Nevertheless, neutrino-induced spallation reactions can change the 
final abundances of some nuclei and we will investigate this in 
future calculations.

\begin{table}
\caption{List of nuclei in reaction network}
\centering
\label{tab:network}
\begin{scriptsize}
\begin{tabular}{cccccc}
\tableline
\tableline
Element & Mass Range & Element & Mass Range & Element & Mass Range \\
\tableline
n & 1 & H & 1--3 & He & 3--6 \\
Li & 6--9 & Be & 7--12 & B & 8--14 \\
C & 9--18 & N & 11--21 & O & 13--22 \\
F & 15--26 & Ne & 17--34 & Na & 17--40 \\
Mg & 20--38 & Al & 21--40 & Si & 22--42 \\
P & 23--48 & S & 24--50 & Cl & 26--42 \\
Ar & 27--44 & K & 29--48 & Ca & 30--50 \\
SC & 32--52 & Ti & 34--54 & V & 36--56 \\
Cr & 38--58 & Mn & 40--62 & Fe & 42--64 \\
Co & 44--66 & Ni & 46--68 & Cu & 48--70 \\
Zn & 51--74 & Ga & 53--86 & Ge & 55--78 \\
As & 57--80 & Se & 59--84 & Br & 61--86 \\
Kr & 64--92 & Rb & 66--92 & Sr & 68--94 \\
Y & 70--96 & Zr & 72--98 & Nb & 74--100 \\
Mo & 77--102 & Tc & 79--104 & Ru & 81--108 \\
 RH & 83--110 & Pd & 86--114 & Ag & 88--116 \\
Cd & 90--118 & In & 92--120 & Sn & 94--126 \\
\tableline
\end{tabular}
\end{scriptsize}
\tablecomments{Nuclear species used in nuclear reaction network for postprocessi
ng.
The mass range given indicates for each element the nuclei with the minimum
and maximum neutron number.}
\end{table}

\begin{table}
\caption{Nuclei for which electron and positron captures are included}
\centering
\label{tab:weaknetwork}
\begin{scriptsize}
\begin{tabular}{cc}
\tableline
\tableline
Mass Number & Nuclides \\
\tableline
1 & n, H \\
21 & Mg, Na, Ne, F, O \\
22 & Mg, Na, Ne, F \\
23 & Al, Mg, Na, Ne, F \\
24 & Si, Al, Mg, Na, Ne \\
25 & Si, Al, Mg, Na, Ne \\
26 & Si, Al, Mg, Na \\
27 & P, Si, Al, Mg, Na \\
28 & S, P, Si, Mg, Na \\
29 & S, P, Si, Al, Mg, Na \\
30 & S, P, Si, Al \\
31 & Cl, S, P, Si, Al \\
32 & Ar, Cl, S, P, Si \\
33 & Ar, Cl, S, P, Si \\
34 & Ar, Cl, S, P, Si \\
35 & K, Ar, Cl, S, P \\
36 & Ca, K, Ar, Cl, S \\
37 & Ca, K, Ar, Cl, S \\
38 & Ca, K, Ar, Cl, S \\
39 & Ca, K, Ar, Cl \\
40 & Ti, Sc, Ca, K, Ar, Cl \\
41 & Ti, Sc, Ca, K, Ar, Cl \\
42 & Ti, Sc, Ca, K, Ar \\
43 & Ti, Sc, Ca, K, Ar, Cl \\
44 & V, Ti, Sc, Ca, K, Ar \\
45 & Cr, V, Ti, Sc, Ca, K \\
46 & Cr, V, Ti, Sc, Ca, K \\
47 & Cr, V, Ti, Sc, Ca, K \\
48 & Cr, V, Ti, Sc, Ca, K \\
49 & Fe, Mn, Cr, V, Ti, Sc, Ca, K \\
50 & Mn, Cr, V, Ti, Sc, Ca \\
51 & Mn, Cr, V, Ti, Sc, Ca \\
52 & Fe, Mn, Cr, V, Ti, Sc \\
53 & Co, Fe, Mn, Cr, V, Ti \\
54 & Co, Fe, Mn, Cr, V, Ti \\
55 & Ni, Co, Fe, Mn, Cr, V, Ti \\
56 & Ni, Co, Fe, Mn, Cr, V, Ti, Sc \\
57 & Zn, Cu, Ni, Co, Fe, Mn, Cr, V \\
58 & Cu, Ni, Co, Fe, Mn, Cr, V, Ti \\
59 & Cu, Ni, Co, Fe, Mn, Cr, V \\
60 & Zn, Cu, Ni, Co, Fe, Mn, Cr, V, Ti \\
61 & Zn, Cu, Ni, Co, Fe \\
62 & Ga, Zn, Cu, Ni, Co, Fe \\
63 & Ga, Zn, Cu, Ni, Co, Fe \\
64 & Ge, Ga, Zn, Cu, Ni, Co, Fe \\
65 & Ge, Ga, Zn, Cu, Ni, Co \\
\tableline
\end{tabular}
\end{scriptsize}
\tablecomments{All nuclei per given mass number for which weak interactions
by \citet{Fuller.Fowler.Newman:1982a,Fuller.Fowler.Newman:1982b} and
by \citet{Langanke.Martinez-Pinedo:2001} are used in the nuclear network.}
\end{table}

\begin{table}
\caption{Nuclei for which neutrino and anti-neutrino captures are included}
\centering
\label{tab:neutrinonetwork}
\begin{scriptsize}
\begin{tabular}{ccc}
\tableline
\tableline
Element & Mass Range & Mass Range \\
 & ($\nu$ capture) & ($\overline{\nu}$ capture) \\
\tableline
n & 1 & 1 \\
H & 1 & 1 \\
He & 6 & 6 \\
Li & 7--9 & 7--9 \\
Be & 8--12 & 8--12 \\
B & 10--14 & 10--14 \\
C & 11--18 & 11--18 \\
N & 13--21 & 13--21 \\
O & 14--22 & 14--22 \\
F & 16--26 & 16--26 \\
Ne & 18--29 & 17--29 \\
Na & 20--32 & 19--22 \\
Mg & 21--35 & 21--35 \\
Al & 22--37 & 22-37 \\
Si & 24--39 & 23--39 \\
P & 26--42 & 25--42 \\
S & 28--42 & 27--42 \\
Cl & 30--42 & 29--42 \\
Ar & 32--44 & 31--44 \\
K & 34--48 & 33--48 \\
Ca & 36--50 & 35--50 \\
Sc & 38--52 &  37--52 \\
Ti & 40--54 & 39--54 \\
V & 42--56 & 41--56 \\
Cr & 44--58 & 43--58 \\
Mn & 45--62 & 45--62 \\
Fe & 48--64 & 46--64 \\
Co & 50--66 & 49--66 \\
Ni & 52--68 & 51--68 \\
Cu & 54--70 & 53--70 \\
Zn & 56--74 & 55--74 \\
Ga & 58--78 & 57--78 \\
Ge & 60--78 & 59--78 \\
Se & 67--84 & 66--84 \\
Br & 69--86 & 68--86 \\
Kr & 71--92 & 70--92 \\
Rb & 73--92 & 72--92 \\
Sr & 77--94 & 74--92 \\
Y  & 79--96 & 78--94 \\
Zr & 81--98 & 80--96 \\
Nb & 83--100 & 82--98 \\
Mo & 85--102 & 84--100 \\
Tc & 87--104 & 86--102 \\
Ru & 89--108 & 88-104 \\
Rh & 91--110 & 90--108 \\
Pd & 94--114 & 92--110 \\
Ag & 96--116 & 95--114 \\
Cd & 98--118 & 97--116 \\
In & 100--120 & 99--119 \\
Sn & ---     & 101--120 \\
\tableline
\end{tabular}
\end{scriptsize}
\tablecomments{Nuclides for which neutrino and anti-neutrino capture
  reactions are included in the nuclear network. The mass range given
  indicates for each element the nucleus with the lowest mass number
  and the nucleus with the highest mass number.}
\end{table}

\begin{figure}[htbp]
  \centering
  \includegraphics[width=\linewidth]{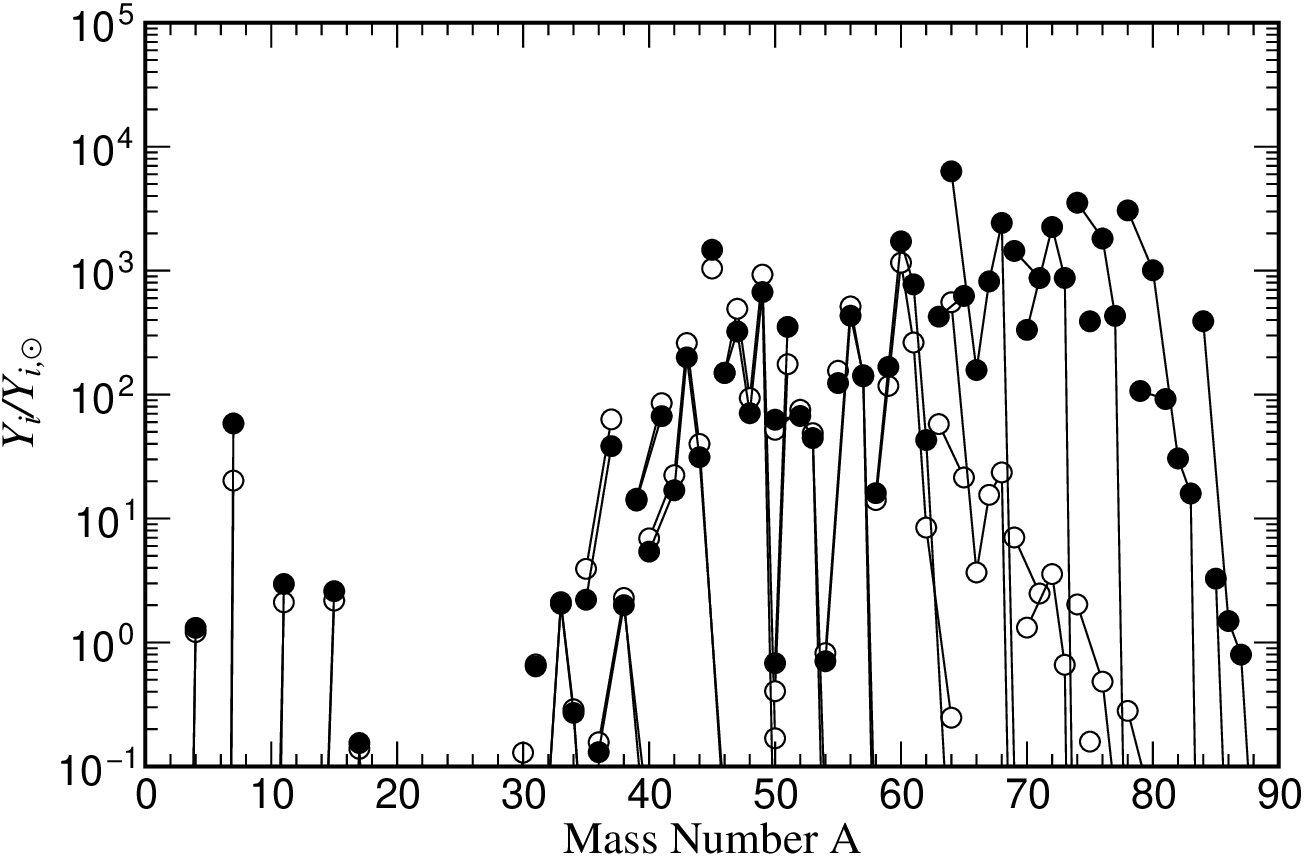}
  \caption{Abundances for model A40 relative to solar abundances
           \citep{Lodders:2003}.
           Two different calculations are shown: with neutrino-induced
           reactions in the network (filled circles) and without
           neutrino-induced reactions in the network (open circles).
           }
  \label{fig:abund-gr040n}
\end{figure}

\begin{figure}[htbp]
  \centering
  \includegraphics[width=\linewidth]{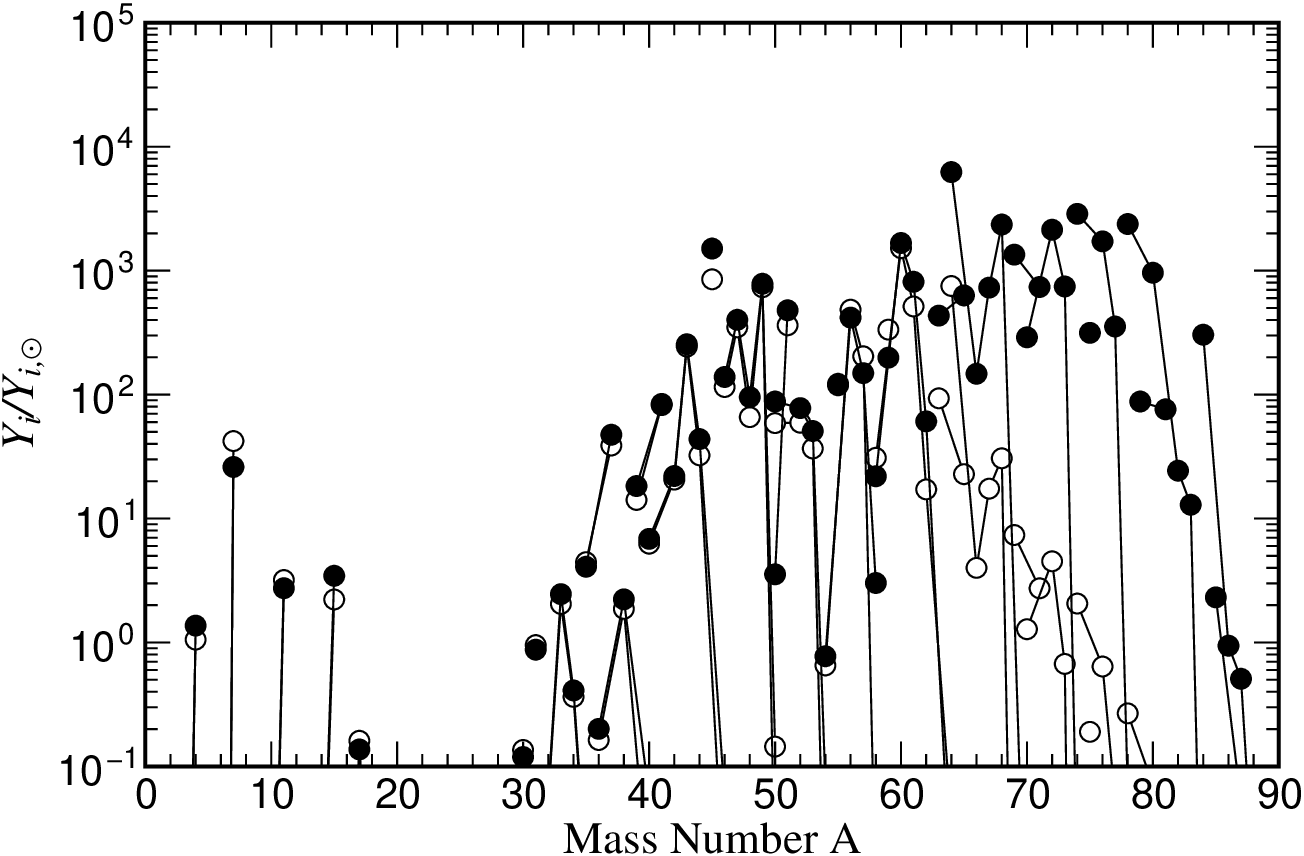}
  \caption{Abundances for model B07 relative to solar abundances
           \citep{Lodders:2003}.
           Two different calculations are shown: with neutrino-induced
           reactions in the network (filled circles) and without
           neutrino-induced reactions in the network (open circles).
           }
  \label{fig:abund-07020n}
\end{figure}

Figure \ref{fig:abund-gr040n} shows the abundances after decay to
stability of 
all nuclei for model A40 integrated over mass
zones with $Y_e>0.5$, including in total $\sim$0.011 M$_{\odot}$.
For these mass zones, we are only concerned with the
Fe-group composition.
In Figure \ref{fig:abund-07020n} integrated abundances after decay to stability
are presented for model B07. In this model, the zones with $Y_e>0.5$
enclose $\sim$0.0066 M$_{\odot}$.
The positions of the mass cut and the explosion energies for both 
models are given in Table \ref{table_runs}.
The isotopic abundances (relative to solar values) result
from postprocessing based on the hydrodynamical profiles and from 
employing the full nuclear reaction network including neutrino and 
antineutrino capture reactions.
Note that unlike earlier supernova nucleosynthesis simulations,
nuclei beyond $A=64$ are also produced 
in appreciable amounts, ranging in fact up to $A=80$ or even beyond,
due to neutrino interactions with matter during the whole period
of explosive processing.
This is discussed in detail in \citet{Frohlich.Martinez.ea:2005}.
These nuclei are mainly produced in the zones close to the mass cut 
where the electron fraction depends strongly on neutrino captures. For
these mass zones relatively high entropies are attained:
$s/k_b \sim$ 30--51.
In nucleosynthesis terms this corresponds to complete explosive Si-burning
with a strong alpha-rich freeze-out which also leaves a finite proton
abundance (0.0007M$_{\odot}$) due to $Y_e$ being larger than 0.5.
The high proton abundance permits the onset of 
an rp-process which, however, does not proceed too far in $A$ as (due to the 
high entropies) the densities are too small.
The abundances result from the accumulation of matter at the waiting-point
nuclei $^{64}$Ge, $^{68}$Se, $^{72}$Kr, $^{76}$Sr. After decay to
stability they produce the high abundances of $^{64}$Zn, $^{68}$Zn,
$^{72}$Ge, and $^{76}$Se. A relatively high abundance of $^{78}$Kr is
also obtained. $^{78}$Kr is considered to be produced by
the p- or $\gamma$-process in the ONe layers of the star. 
Chemical evolution studies \citep[e.g.][]{Timmes.Woosley.Weaver:1995} 
underproduce $^{64}$Zn by about a factor 5. 
A possible site for the production of $^{64}$Zn is the modest early-time
neutrino-driven wind occurring after core bounce in
supernovae \citep{Woosley.Hoffman:1992}. 
\citet{Umeda.Nomoto:2005} have found that the
$^{64}$Zn/$^{56}$Fe ratio is enhanced \emph{if} Ye is close to 0.5
and the explosion energy is as high as $\sim 10^{52}$ erg.
Our proton-rich environment constitutes an alternative or
complementary production site.

For the intermediate mass elements the main improvement compared to
earlier calculations is the higher production of individual nuclei like
$^{45}$Sc and $^{49}$Ti.
Scandium is mainly produced by the $\beta^+$-decays originating from 
$^{45}$Cr and $^{45}$V decaying via $^{45}$Ti to $^{45}$Sc. 
Different calculations of abundance yields 
(TNH96,  
\citeauthor{Woosley.Weaver:1995} \citeyear{Woosley.Weaver:1995},
\citeauthor{Chieffi.Limongi:2002a} \citeyear{Chieffi.Limongi:2002a})
fail to predict the observed abundance of
scandium \citep{Gratton.Sneden:1991,Cayrel.Depagne.ea:2004}. 
Our calculations show that Sc
can be consistently produced with iron in the inner regions of the
supernova where $Y_e$ is higher than 0.5. The ejected yield
of Sc is $10^{-6}$~M$_\odot$ which
is a factor of 10 larger than the value obtained for a similar star
by TNH96.
If we assume that our
total production of Fe is similar to the one obtained in 
TNH96  our Sc yield will be consistent with observations. $^{49}$Ti is
underproduced by a factor 5 in the chemical evolution studies
of~\citet{Timmes.Woosley.Weaver:1995}. 
The nucleus $\mathrm{^{49}Ti}$ originates from the decay chain of
$\mathrm{^{49}Mn}$ which decays via $^{49}$Cr and $^{49}$V to $^{49}$Ti. 
After decay to stability, 
the resulting yield of $\mathrm{^{49}Ti}$
is $\sim5\times 10^{-6}$ M$_\odot$.
We find that the origin of the differences in nucleosynthesis yields is
a consequence of an electron fraction above 0.5 which is due to a 
consistent treatment of all weak interaction processes on free nucleons.
The obtained $Y_e$ values are not sensitive to the inclusion of
neutrino and antineutrino captures on nuclei.

\begin{figure}[htbp]
  \centering
  \includegraphics[width=\linewidth]{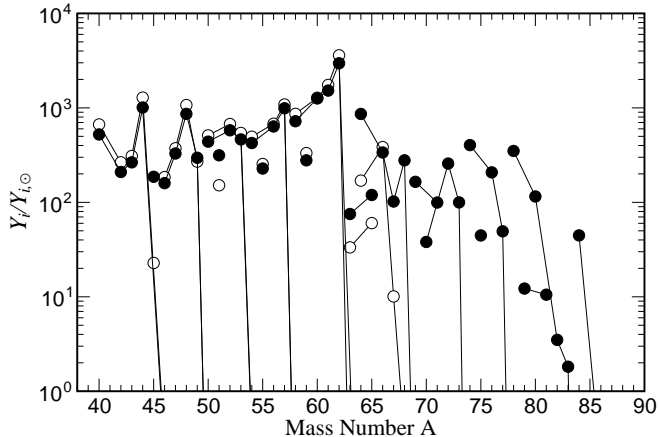}
  \caption{Combined abundances of this work (model A40) and
           TNH96. The open circles
           are the combined abundances and the filled circles are the original
           abundances of the above reference.
           For details see text.
           }
  \label{fig:abund-cffkt}
\end{figure}

\begin{figure}[htbp]
  \centering
  \includegraphics[width=\linewidth]{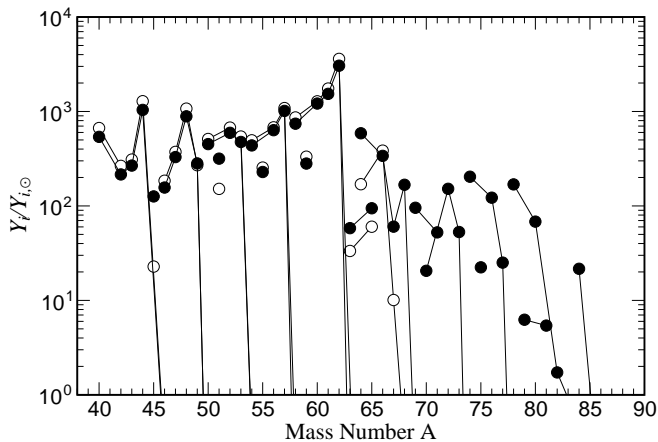}
  \caption{Combined abundances of this work (model B07) and
           TNH96. The open circles
           are the combined abundances and the filled circles are the original
           abundances of the above reference.
           For details see text.
           }
  \label{fig:abund-cffkt2}
\end{figure}

In the absence of a (yet) complete nucleosynthesis calculation covering the
entire region responsible for
Fe-group production we combine our abundances with the results of 
TNH96 (see Figures \ref{fig:abund-cffkt} and \ref{fig:abund-cffkt2}).
The two calculations are combined in such a way that the resulting amount
of Fe-group elements is the same as in this earlier work.
For the inner zones, where neutrino and antineutrino capture reactions play 
an important role, the results of the present calculation are used. They 
constitute about 30\% of the total production of Fe-group elements.
For the other zones where neutrino/antineutrino captures 
do not have significant influence on the final $Y_e$
we use the abundance results from TNH96.
This procedure allows us to estimate the influence of a consistent treatment
of weak interaction processes on the total production of Fe-group elements.
To further solidify these results full nucleosynthesis calculations are being
performed based on the exploding models.

In a recent work, \citet{Pruet.Woosley.ea:2004} have studied 
a similar scenario (with a similar philosophy) for the innermost
ejected layers, based on a 2D simulation by \citet{Janka.Buras.Rampp:2003}.
While the original 2D simulation did not yield an explosion, 
omitting the velocity-dependent terms from the neutrino momentum equation
resulted in a successful explosion.  They find results in agreement with
our results. However, they do not report on the production of heavy nuclei 
with $A>64$
due to the fact that they utilize the $Y_e$ from the hydrodynamic
calculation but do not include neutrino interactions in their network.

\begin{figure}
  \centering
  \includegraphics[width=\linewidth]{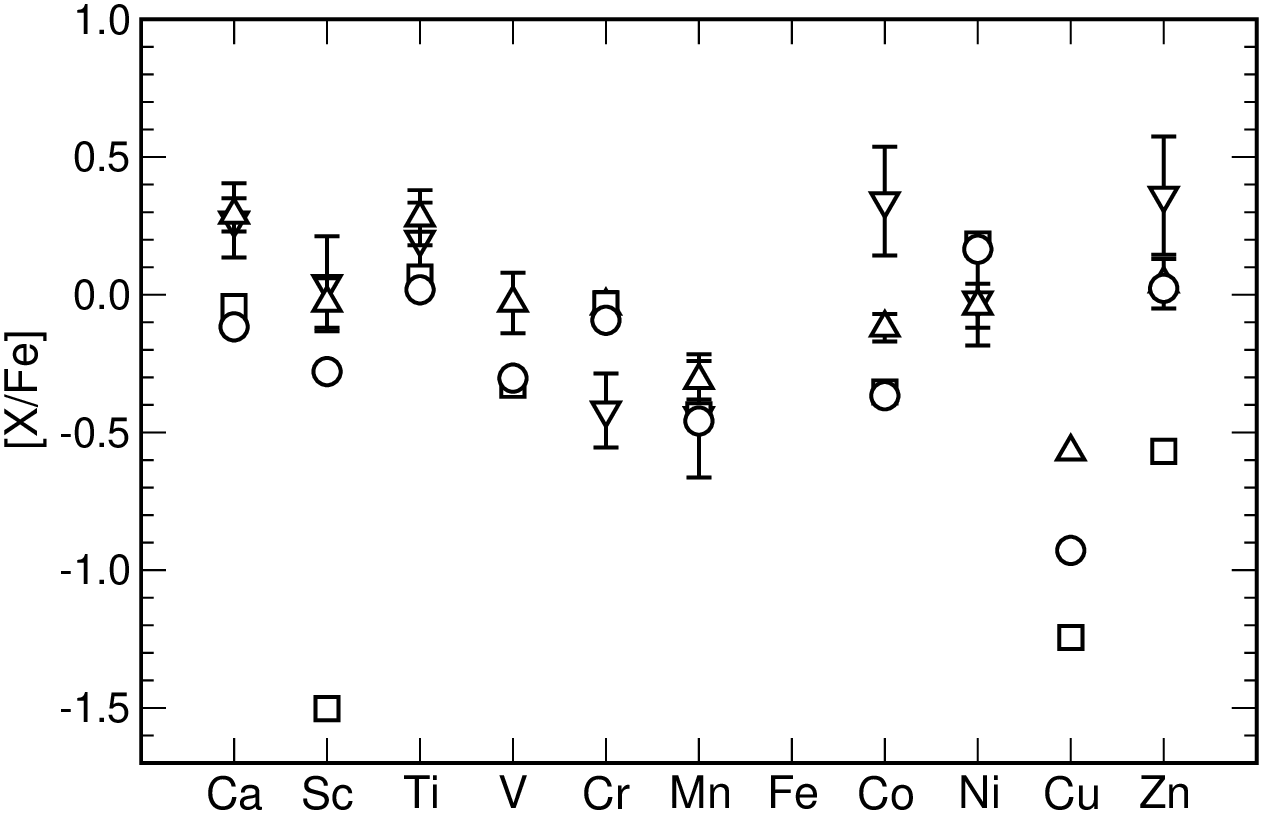}
  \caption{Comparison of elemental overabundance in the mass range 
           Ca to Zn for different calculations.
           The triangles with error bars represent observational data.
           The triangles facing upwards \citep{Gratton.Sneden:1991} originate from 
           an analysis of stars with $-2.7<[\mathrm{Fe/H}]<-0.8$.
           The traingles facing downwards \citep{Cayrel.Depagne.ea:2004} is 
           data for a sample of  extremely metal poor stars 
           ($-4.1 <[\mathrm{Fe/H}]< -2.7$).
           The circles are abundances of this work combined with 
           the work of TNH96 to obtain
           the same amount of Fe-group elements. The squares show 
           the pure abundances of the previous reference.
          }
  \label{fig:compare}
\end{figure}

Figure \ref{fig:compare} shows elemental abundances of two
calculations compared to two sets of observational data. One set 
of observational data \citep{Gratton.Sneden:1991} originates from 
a sample of stars with $-2.7<[\mathrm{Fe/H}]<-0.8$, relevant for the average
type II supernova contribution. The second set of observational data represents 
a sample of extremely metal-poor stars \citep{Cayrel.Depagne.ea:2004}.
The nucleosynthesis results of this work are shown in combination with the 
results of the earlier calculation as shown in Figure \ref{fig:abund-cffkt}
(the results corresponding to Figure \ref{fig:abund-cffkt2} are very similar 
with the exception of Zn and heavier nuclei).
For comparison, the theoretical prediction by TNH96 is also shown.
We see clearly an improvement for Sc and the heavy elements Cu and Zn.

\section{Conclusions}
\label{sec:conclusions}

Presently, self-consistent core collapse supernova simulations in 1D 
do not lead to successful explosions while 2D models show some promise. 
Remaining uncertainties in neutrino
opacities and/or the expected strong influence of convection (due to
hydrodynamic instabilities caused by entropy gradients and/or rotation and
magnetic fields) are likely to change this result.
They may lead either to higher
neutrino luminosities or higher efficiencies of neutrino energy deposition via
neutrino and antineutrino captures on nucleons. In order to examine the
accompanied nucleosynthesis 
in 1D calculations of successful
explosions, we performed simulations with variations in neutrino
scattering cross sections on nucleons and/or neutrino and antineutrino
captures on neutrons and protons. In both cases successful explosions
emerge with an interesting evolution of the $Y_e$ gradient in the innermost
ejecta, which were followed up by a postprocessing for nucleosynthesis purposes.

The detailed nucleosynthesis calculations with a consistent treatment of
all weak interactions show an electron fraction $Y_e>0.5$, i.e.\ 
a slightly proton-rich environment with relatively high entropies of up to
$\sim$50 $k_B$ per nucleon. This causes complete Si-burning with an alpha-rich (and
proton-rich) freeze-out. About 0.0007 M$_\odot$ of hydrogen
remain in the innermost ejecta 
and do not stem from mixing this
matter in from the hydrogen envelope. Such a proton-rich environment at
relatively high entropies permits to produce
also nuclei beyond $A$=64, up to $A$=80, with a
major contribution to $^{64}$Zn.
The rp-like process does not extend to masses beyond A=80--100 as the 
high entropies imply
too small densities for a path at very small proton separation energies.

In addition, we find improvements within the Fe-group. The strong overabundances
of $^{58,62}$Ni found in previous (too neutron-rich) environments are
reduced. $^{45}$Sc and $^{49}$Ti are enhanced to permit predictions
closer to solar proportions. Especially the emergence of $^{45}$Sc seems to
be a solution to the previously not understood abundance of this only
stable isotope of Sc.
This discussion is also of interest
with respect to $^{44}$Ti, made in 
the alpha-rich freeze-out in the inner explosive ejecta. $^{44}$Ti is 
sensitive to  $Y_e$ and reduced in the mass range where $^{45}$Sc is high. 
This will influence the overall predictions of $^{44}$Ti.

Values of $Y_e>0.5$ are due to the neutrino interactions with matter 
under electron non-degenerate conditions in a convectively unstable domain and
thus related to the explosion mechanism. The effect of neutrinos
decreases with $1/r^2$ and about 50-60\% of the Fe-group ejecta (the
outer part of explosive, complete Si-burning) is determined by values of
$Y_e$ equal to or close to the initial values inherited from stellar
evolution. In this first study we have tried to give an estimate for the
overall Fe-group composition based on such a superposition of the present
results for the innermost ejecta with those of TNH96.
Future investigations will require to perform full nucleosynthesis calculation
for complete stars based on these exploding models. They will also require
a sensitivity test of the nucleosynthesis results 
to the scaling factors for neutrino-induced reactions
discussed in Figure \ref{fig:evol} and in combination with the position of the
mass cut and the explosion energy. This should be considered in order to 
reproduce results for supernovae where detailed observational information 
in abundances, gamma-ray emitters and explosion energies is available.

\acknowledgements{This work is partly supported by the Swiss SNF grant 
200020-105328
and by the Spanish MCyT and European Union ERDF under contracts
AYA2002-04094-C03-02 and AYA2003-06128.
The neutrino transport calculations have been performed on the CITA Itanium I.
The work has been partly supported by the United States National 
Science Foundation under contract PHY-0244783, by the United States 
Department of Energy, through the Scientic Discovery through Advanced 
Computing Program.  Oak Ridge National Laboratory is managed by 
UT-Battelle, LLC, for the U.S. Department of Energy under contract 
DE-AC05-00OR22725.
A.M. is supported at the Oak Ridge National Laboratory, which is
managed by UT-Battelle, LLC for the DOE under contract DE-AC05-00OR22725.
He is also supported in part by a DOE ONP Scientific Discovery through
Advanced Computing Program grant.
}


\end{document}